\documentclass[prl,aps,preprint,showpacs]{revtex4}
\usepackage{graphicx}

\usepackage{epstopdf}
\begin{document}

\title{Critical Velocities for Roton and Super-Flow Quantum Turbulence in Liquid $^4$He} 
\author{V. I. Kruglov}
\address{Physics Department, The University of Auckland, Private Bag
         92019, Auckland, New Zealand}

\begin{abstract}
Two different types of transitions of the superfluid $^4$He to quantum turbulence regimes are studied for $1{\rm D}$ geometry in the case when the influence of the normal fluid on superfluid flow is suppressed. It is shown that the roton mechanism of transition to quantum turbulence leads to a critical velocity satisfying the relation $v_c\propto d^{-1/4}$. 
In the super-flow mechanism, the transition to quantum turbulence arises when the ``quantum Reynolds number" is about $10^3$ and the critical velocity depends on channel size $d$ as $v_c\propto d^{-1}$ in agreement with the equations of motion for a superfluid component of the liquid $^4$He being disturbed by small fluctuations of the normal fluid.

\end{abstract}


\maketitle

The understanding of the existence of critical velocity in superfluid $^4$He and its microscopic nature is a long-standing problem which also closely related to superfluid turbulence \cite{Vin,Vinen}. The critical velocity in a superfluid is the threshold above which the flow of the superfluid component becomes dissipative and the property of superfluidity is lost. The first well known criterion for critical velocity, $v_c={\rm min}~\varepsilon (p)/p$  where $\varepsilon (p)$ is the energy spectrum of the elementary excitations at momentum $p$, was proposed by Landau \cite{Kha}. This criterion yields
the value $v_c\simeq 60~m/s$ which does not depend on capillary parameters.
The Feynman criterion \cite{Fey}, following Onsager, is based on the quantization of superfluid circulation and defines the velocity $v_c\simeq(\hbar/md){\rm ln}(d/a)$ at which the vortices can be excited. Here $d$ is the channel size and $a\simeq 4$\AA~ is the vortex core radius. These two different criteria can differ greatly from experimental data \cite{Eric}; however, the Feynman estimation is much closer to experimental values.
The difficulty in this problem also arises due to the existence of different breakdown mechanisms of the superfluidity \cite{Eric,Schwar,Nore}. Among the several problems associated with critical velocity the type of elementary excitation which would be responsible for the onset of dissipation and for a critical velocity in the superfluid plays an important role. In Feynman's approach such elementary excitations are vortices excited in the superflow of the channel \cite{Fey}. 

We study in this letter the existence of the critical velocity in superfluid $^4$He when the normal fluid does not influence the superfluid flow. The most efficient method of preventing the normal fluid from interfering with the flow is by introducing superleaks at the ends of the capillary \cite{Craig}. The narrow paths in the fine powder act as solid walls to the normal fluid and prevent it from entering the flow region. As soon as the normal component is sufficiently suppressed by application of superleaks in the He II flow path between parallel plates or in a capillary it is found that the experimental results are well described by the temperature independent critical velocity $v_c\propto d^{-1/4}$  \cite{Ver,VanAl,Van,Bruyn,Putt} when $T<2~{\rm K}$ and the separation of the plates is $d_o<d<d_c$ where $d_o=0.8\cdot 10^{-6}~{\rm cm}$ and $d_c=1~{\rm cm}$.

Our explanation of this relation for $v_c$ is based on the cluster model of the roton \cite{Krug,Kru} in liquid $^4$He. It was shown in \cite{Krug,Kru} that the roton is an excitation of a cluster of $13$ atoms having a central atom surrounded by a shell of $12$ atoms situated at the vertices of a regular icosahedron, although in this picture each atom can deviate about its location in the cluster. We show that rotational excitations of the roton cluster of $13$ atoms in the superflow are the elementary excitations responsible for the onset of dissipation and hence for a critical velocity in the superfluid for a range of capillary widths $d$.

The elementary rotational excitations are excited by collision with the surface of the channel. The rotons can exchange between each other by the rotational energy in the superfluid and when the velocity of the flow becomes greater of some critical value $v_c$ the number of the elementary rotational excitations grows in the form of an avalanche and 
 the superfluid motion breaks down because the irrotational superfluidity requirement ${\rm curl}{\bf v_s}=0$ for macroscopic superfluid velocity is no longer satisfied. In this microscopic picture we assume that interaction of these rotational excitations forms the vortices in the superflow of the channel. It is shown below by dimensional analysis that the roton mechanism of transition to quantum turbulence leads to a critical velocity given by $v_c\propto d^{-1/4}$, in agreement with experimental data \cite{Ver,VanAl,Van,Bruyn,Putt}. 

We also consider in the paper another type of turbulence (super-flow quantum turbulence) which is  quasi-classical and arises when the size between parallel plates $d>1~{\rm cm}$ and the dimensionless parameter ${\cal R}=d/\lambda$ is about $10^3$ where $\lambda=2\pi\hbar/mv$ is the characteristic de Broglie wavelength of helium atoms in the channel flow.
It is found that a super-flow quantum turbulence arises when $v>v_c$ where the critical velocity satisfies the relation $v_c\propto d^{-1}$. This type of turbulence differs considerably from the quantum turbulence when the normal fluid is not suppressed \cite{Vin,Vinen} and, in particular, the critical velocity by Feynman criterion in this case leads to a value about three orders smaller than was found in the experiments \cite{Putt} at $d\simeq1~{\rm cm}$ for normal mass density $\rho_n\ll \rho$.

The Schr\"odinger-type equation for the wavefunction $\Psi(a_1,a_2,a_3,t)$ describing the roton cluster in the space of ellipsoidal parameters $a_k~(k=1,2,3)$ as a complete system of $N_c$ bound atoms is \cite{Krug,Kru}: 
\begin{equation}
 i\hbar\frac{\partial\Psi}{\partial t}=\left(-\frac{\hbar^{2}}{2M_0}
 \sum_{k=1}^{3}\frac{\partial^{2}}{\partial a_{k}^{2}}+\frac{M_0}{2}
 \sum_{k=1}^{3}\omega_{k}^{2}a_{k}^{2}+\frac{{\cal G}}{a_{1}a_{2}a_{3}}
 \right)\Psi,
\label{1}
\end{equation}
where $M_0$ is the effective mass of the roton cluster and ${\cal G}$ is an interaction constant given by
\begin{equation}
M_0=\frac{N_c m}{7},~~~{\cal G}=\frac{15a_{0}\hbar^{2}N_{c}(N_c-1)}{7m}.
\label{2} 
\end{equation}
For spherically symmetric roton states ($\bar{a}_k=\bar{a}$) the frequency $\omega=\omega_k$ ($k=1,2,3$) is  
\begin{equation}
\omega=\frac{\hbar}{m\bar{a}^2}\sqrt{\frac{15 (N_c-1) a_0}{\bar{a}}},
\label{3}
\end{equation}
where $\bar{a}$ is the effective radius of the roton cluster and $a_0=2.2~$\AA~is the s-scattering wavelength of $^4$He atoms:
\begin{equation}
\bar{a}=\left(\frac{3N_cm}{4\pi \rho}\right)^{1/3},~~~a_0=\frac{m}{\hbar^2}\int_{a_0}^\infty U(r) r^2dr.
\label{4}
\end{equation}
Here $\rho$ is the mass density of liquid helium and the s-scattering wavelength is calculated by intermolecular Lennard-Jones potential $U(r)=4\epsilon[\left(r_0/r\right)^{12}-\left(r_0/r\right)^6]$ for $^4$He. The stability condition of the roton cluster has the form $\lambda_D\geq 2\bar{a}$ where $\lambda_D=2\pi\hbar/\sqrt{3mk_BT}$ is the thermal wavelength. Using the quadratic approximation in the potential of Eq. (\ref{1}) the eigenenergies \cite{Krug,Kru} to good accuracy are
\begin{equation}
E_{n_{1}n_{2}n_{3}}=V_{0}+\sum_{k=1}^{3}\hbar\Omega_{k}(n_{k}+1/2).
\label{5} 
\end{equation}
Here $n_k=0,1,2,...$ (at $k=1,2,3$) are the quantum numbers of a 3D quantum harmonic oscillator. In the spherically symmetric case Eq. (\ref{1}) yields the eigenfrequencies $\Omega_{k}$ as
\begin{equation}
\Omega_1=\Omega_2=\sqrt{2}\omega,~~~\Omega_3=\sqrt{5}\omega.  
\label{6}
\end{equation}
In the general case rotational quantum states of the roton clusters can also be excited. If an inequality $\xi=B_c/\hbar\omega\ll 1$ is satisfied where $B_c=\hbar^2/2 I_c$ is the rotational constant  and $I_c$ is the moment of inertia of the roton cluster, then for small vibrational quantum numbers $n_k$ and small angular momentum quantum numbers $J$ the eigenenergies of the roton are 
\begin{equation}
E_{n_{1}n_{2}n_{3}}^J=V_{0}+\sum_{k=1}^{3}\hbar\Omega_{k}(n_{k}+1/2)+B_cJ(J+1).
\label{7} 
\end{equation}
This equation comes from standard quantum mechanical methods in Eq.(\ref{1}) as developed for rotational-vibrational molecular spectra.  
In this general case the number of atoms $N_c$ in the roton cluster satisfies the condition \cite{Kru}
\begin{equation}
E_{n_{1}n_{2}n_{3}}^J-E_{000}^0=\Delta,~~~\Delta=\epsilon-\epsilon_0,
\label{8} 
\end{equation}
where $\Delta$ is the roton gap defined by the parameters $\epsilon=-U(r_m)$ and $\epsilon_0=-U(2a_0)$. Here $r_m=2^{1/6}r_0$~is the distance for which
the intermolecular potential energy $U(r)$ has a minimum and $2a_0$ is the mean distance between helium atoms in the liquid $^4$He. A self-consistent-field Hartree-Fock method \cite{Ahl,Aziz} yields the parameters: $\epsilon/k_B=10.6~{\rm K}$ and $r_m=2.98~$\AA, hence from Eq.(\ref{8}) it follows that $\Delta/k_B=8.65~{\rm K}$. Using Eqs. (\ref{6}-\ref{8}) the equation for the number $N_c$ in the roton cluster is 
\begin{equation}
\hbar\omega\left[\sqrt{2}(n_1+n_2)+\sqrt{5}n_3\right]+B_cJ(J+1)=\Delta.
\label{9} 
\end{equation}
Because we assume $\xi\ll 1$, the term $B_cJ(J+1)$ can be neglected and from Eqs. (\ref{3},\ref{4},\ref{9}) follows that the integer number $N_c$ is given by
\begin{equation}
N_c^5-\alpha[\sqrt{2}(n_1+n_2)+\sqrt{5}n_3]^6(N_c-1)^3=0,
\label{10} 
\end{equation}
where $\alpha=(\Delta_0/\Delta)^6$, and the parameter $\Delta_0$ is 
\begin{equation}
\Delta_0=\frac{\sqrt{15}\hbar^2}{ma_0^2}\left(\frac{a_0}{q_0}\right)^{5/2},~~~q_0=\left(\frac{3m}{4\pi \rho}\right)^{1/3}.
\label{11} 
\end{equation}
The smallest solution to Eq. (\ref{10}) is $N_c=13.1$ \cite{Kru} in a symmetric vibrational state with $n_1=n_2=0$ and $n_3=1$, hence the most stable roton clusters should be those consisting of $13$ helium atoms. We note that for roton clusters with $N_c=13$, $\xi={\cal O}(10^{-2})$ and $\bar{a}=5.22$~\AA.

The kinetic energy flow density of the superfluid between parallel plates for the case when there is no normal fluid in superfluid flow ($\rho_n=0$) is
\begin{equation}
F\equiv\frac{1}{S}\frac{\delta E_s}{\delta t}=\frac{1}{2}\rho_s v_s^3=\frac{1}{2}\rho v_s^3,
\label{12} 
\end{equation}
where $S=Ld$ is the area of the cross section and $\rho=\rho_s$. Let us consider $1{\rm D}$ geometry (the width of the plates $L$ is much greater then the distance $d$ between plates); then the kinetic energy flow per unit length along the width of the parallel plates  is given by $f=Fd$ or in the limit $L\rightarrow\infty$:
\begin{equation}
f\equiv \lim_{L\rightarrow\infty}\frac{1}{L}\frac{\delta E_s}{\delta t}=\frac{1}{2}\rho v_s^3d.
\label{13} 
\end{equation}
The function $f$ has such a form only for $v_s\leq v_c$ because when the velocity of the flow is greater then $v_c$ the onset of dissipation occurs in the liquid $^4$He. Hence the critical value of $f$ given by $f_c=\rho v_c^3d/2$ can be treated as a threshold parameter of the superfluid. Another important parameter of the superfluid is the minimal rotational energy $\varepsilon_r=E_{001}^1-E_{001}^0$ which rotons can exchange in the superfluid between the rotational states with angular momentum quantum numbers $J=1$ and $J=0$. From Eq. (\ref{7}) it follows that this parameter is defined by the moment of inertia of the roton cluster $I_c$ as $\varepsilon_r=2B_c=\hbar^2/I_c$.

We may assume that when the normal fluid is suppressed the critical velocity depends on three parameters: the threshold parameter $f_c$, the minimal rotational energy $\varepsilon_r$ and the minimal angular momentum $\hbar$ which rotons can exchange in the superfluid. The parameters $\varepsilon_r$ and $\hbar$ are connected with the dissipation process of the energy and angular momentum in the liquid $^4$He at $v_s=v_c$. Thus it is assumed that $v_c=Q(f_c,\varepsilon_r,\hbar)$ where $Q$ is some unknown function of three variables. Taking into account that the threshold parameter has the form $f_c=\rho v_c^3d/2$ the relation for critical velocity can be rewritten in an equivalent form 
\begin{equation}
v_c={\cal Q}(\rho d,\varepsilon_r,\hbar),
\label{14} 
\end{equation}
where ${\cal Q}$ is a new unknown function of three independent variables $\rho d$, $\varepsilon_r$ and $\hbar$. It can be shown by dimensional analysis that there exists a unique function ${\cal Q}$ of the variables $\rho d$, $\varepsilon_r$ and $\hbar$ which has the dimensionality ${\rm cm}~{\rm s}^{-1}$. This function has the form ${\cal Q}=\gamma(\rho d)^n\varepsilon_r^k\hbar^l$ where $\gamma$ is a dimensionless parameter and the unique powers are: $n=-1/4$, $k=3/4$ and $l=-1/2$. Hence, the general relation given by Eq. (\ref{14}) yields the unique equation
\begin{equation}
v_c=\Gamma d^{-1/4},
\label{15} 
\end{equation}
where the constant parameter $\Gamma$ has an explicit form:
\begin{equation}
\Gamma=\gamma\left(\frac{\varepsilon_r^3}{\rho\hbar^2} \right)^{1/4}=\frac{\gamma\hbar}{\rho^{1/4}I_c^{3/4}}.
\label{16} 
\end{equation}
We emphasize that the Eq. (\ref{15}) coincides with the known empirical relation \cite{Ver,VanAl,Van,Bruyn,Putt} for critical velocity in the case when the normal component is sufficiently suppressed by application of superleaks. Moreover, the explicit expression given by the Eq. (\ref{16}) allows us to estimate the constant parameter $\Gamma$. It is assumed in dimensional analysis \cite{Bar} that the dimensionless parameter (in our case $\gamma$) has a magnitude of order unity. More exactly, one may guess that 
${\cal O}(10^{-1})\leq\gamma\leq{\cal O}(10)$. The evaluation of the constant parameter $\Gamma$ by Eq. (\ref{16}) for roton clusters with $N_c=13$ and $\gamma={\cal O}(10^{-1})$ yields $\Gamma={\cal O}(1)~{\rm cm}^{5/4}{\rm s}^{-1}$ which is in agreement with the experimental value $\Gamma=1~{\rm cm}^{5/4}{\rm s}^{-1}$ in Ref. \cite{Ver,VanAl,Van,Bruyn,Putt}. Because there is no other characteristic energy in liquid $^4$He comparable to $\varepsilon_r$ in size, this estimation demonstrates that roton breakdown mechanism of the superfluidity explains the experimental relation $v_c=\Gamma d^{-1/4}$ for critical velocities when $d_o<d<d_c$. 
We also emphasize that the Eqs. (\ref{15},{16}) shows that the superfluidity is a quantum phenomenon because $v_c\rightarrow 0$ in the classical limit $\hbar\rightarrow 0$.

The experiment also demonstrated the temperature dependence of the critical velocity \cite{Putt} showing that $v_c$ dropped suddenly to zero above about temperature $T_0=2~{\rm K}$ and below this temperature $v_c$ was constant. We may estimate the characteristic temperature $T_0$  by condition $\lambda_D> 2d_H$ where $d_H=2a_0=4.4~$\AA~is the effective diameter of helium atoms. 
This condition means that the superfluidity takes place only for temperatures $T< T_0$ and leads to characteristic temperature:
$T_0=\pi^2\hbar^2/12mk_Ba_0^2$.
This yields the temperature $T_0=2.05~{\rm K}$ which coincides with the experimental characteristic temperature. 

We assume that for thicknesses of the channel $d\geq d_c$ where $d_c$ is the critical value for existing of the relation $v_c\propto d^{-1/4}$, the physical breakdown mechanism of the superfluidity is different and it is connected with the characteristic interaction lengths $l>2\bar{a}=10.44$~\AA. In this case the wavenumbers belongs in  the phonon region $k<k_0$ where $k_0=\pi/\bar{a}=0.6$~\AA$^{-1}$. It can be shown that in this region of wavenumbers the effective pair interacting potential has the form (which will be published elsewhere): $\tilde{U}(r)=U(r)$ for $r\in[a_0,a_m]$ and $\tilde{U}(r)=0$ for $r\notin[a_0,a_m]$ where $U(r)$ is the Lennard-Jones potential \cite{Ahl,Aziz} and $a_m=2r_m-a_0=3.76$~\AA. This means that in the effective Hamiltonian for liquid $^4$He there is a cut-off in the intermolecular potential for distances $r<a_0$ and $r>a_m$. The first cut-off $r<a_0$ is related with the repulsive part of the intermolecular potential in the modified Born approximation \cite{Kru} (see Eq. (\ref{4})). The cut-off in the region $r>a_m$, where $a_m$ is defined by relation $r_m-a_0=a_m-r_m$, is connected with the long-range attractive forces between one atom and atoms in the surrounding bulk
liquid summing to zero. 

The many-body effective Hamiltonian for the region of wavenumbers $k<k_0$ yields the Heisenberg equation of motion for the field operator as

\begin{figure}
\centering
\includegraphics[width=0.8\textwidth]{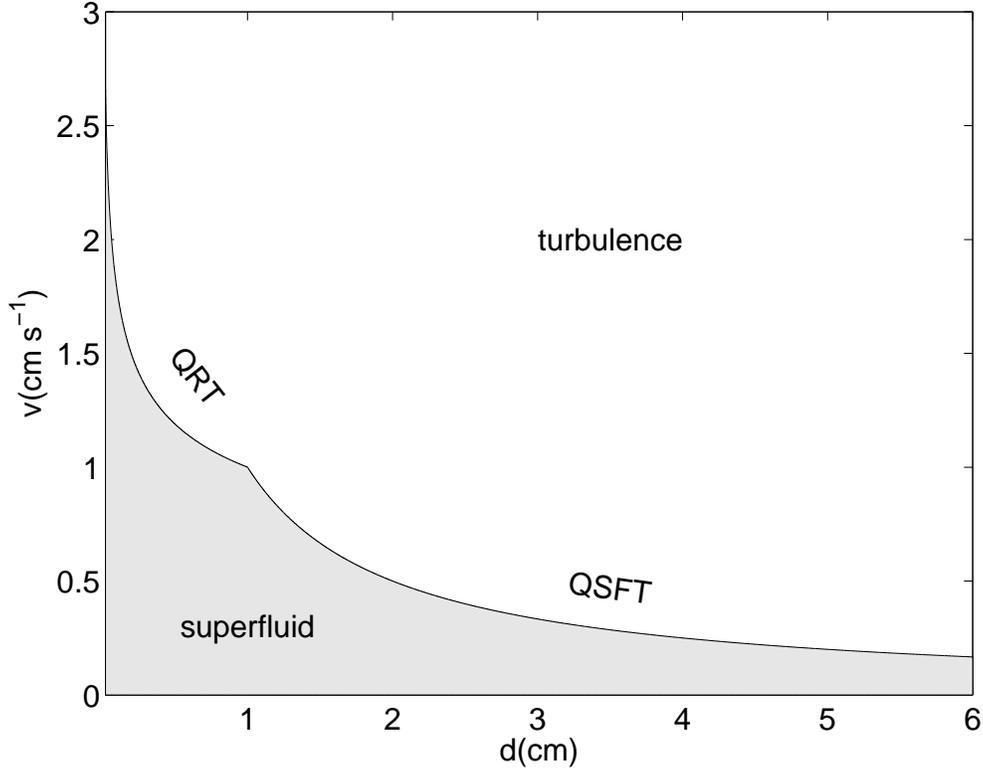}
\caption{Flow velocity $v$ for $^4$He in a channel of width $d$ showing
regions of superfluidity and turbulence.  This is for 1D geometry with
$T<2$\,K and the influence of the normal fluid on superfluid flow suppressed
by superleaks.  The line shows the critical velocity $v_c$ which separates
superfluidity from quantum rotational turbulence for widths $d<1$\,cm (Eqs. (\ref{15},{16})) and from quantum super-flow turbuluence for widths $d>1$\,cm (Eqs. (\ref{22},{23})).}
\label{fig:1}
\end{figure}

\[
i\hbar\frac{\partial\hat{\psi}}{\partial t}=-\frac{\hbar^2}{2m}\nabla^2\hat{\psi}+G\hat{\psi}^{\dagger}\hat{\psi}\hat{\psi},
\]
where the coupling constant $G=4\pi\int_{0}^{\infty}\tilde{U}(r)r^2dr=1.79\cdot 10^{-37}~{\rm g}~{\rm cm}^{5}~{\rm s}^{-2}$.  We note that the parameter $G$ significantly differs from the coupling parameter $g=4\pi a_0\hbar^2/m=4.63\cdot 10^{-38}~{\rm g}~{\rm cm}^{5}~{\rm s}^{-2}$ in Gross-Pitaevskii equation for dilute helium gas \cite{Pit}. Using the field operator in the form $ \hat{\psi}({\bf r},t)=\psi_s({\bf r},t)+\hat{\psi}_n({\bf r},t)$ where $\langle\hat{\psi}({\bf r},t)\rangle=\psi_s({\bf r},t)$ and
$\langle\hat{\psi}_n({\bf r},t)\rangle=0$ we may write the full mass density of the liquid helium as the sum $\rho=\rho_s+\rho_n$ with
\[
\rho_s({\bf r},t)=m|\psi_s({\bf r},t)|^2,~~~\rho_n({\bf r},t)=m\langle\hat{\psi}_n^{\dagger}({\bf r},t)\hat{\psi}_n({\bf r},t)\rangle.
\]
We consider the case $\rho_n\ll \rho_s$; then in the dynamic Popov approximation \cite{Min} the anomalous density $\langle\hat{\psi}_n\hat{\psi}_n\rangle$  and the three-field correlation function $\langle\hat{\psi}_n^{\dagger}\hat{\psi}_n\hat{\psi}_n\rangle$ can be neglected in an equation for macroscopic wavefunction $\psi_s({\bf r},t)$: 
\begin{equation}
i\hbar\frac{\partial\psi_s}{\partial t}=-\frac{\hbar^2}{2m}\nabla^2\psi_s+G|\psi_s|^2\psi_s+2G\langle\hat{\psi}_n^{\dagger}\hat{\psi}_n\rangle\psi_s.
\label{17} 
\end{equation}
The Heisenberg equation also yields the linearized equation for the field operator $\hat{\psi}_n({\bf r},t)$ as
\begin{equation}
i\hbar\frac{\partial\hat{\psi}_n}{\partial t}=-\frac{\hbar^2}{2m}\nabla^2\hat{\psi}_n+2G|\psi_s|^2\hat{\psi}_n+G\psi_s^2\hat{\psi}_n^{\dagger}.
\label{18} 
\end{equation}
The homogeneous solution of Eq. (\ref{17}) at $\rho_n\ll \rho_s$ is $\psi_{s}^h=\psi_{s0}\exp(-i\mu t/\hbar)$ where $\mu=G|\psi_{s0}|^2$ is the effective chemical potential and $\psi_{s0}=\sqrt{\rho_{s0}/m}\exp(i\Phi_0)$ is the constant amplitude. 
In this case the solution of Eq. (\ref{18}) for the field operator $\hat{\psi}_n({\bf r},t)$ is
\[
\hat{\psi}_n({\bf r},t)= \frac{e^{-i\mu t/\hbar}}{\sqrt{V}}\sum_{{\bf k}}\frac{\left(\hat{c}_{{\bf k}}e^{-i\omega(k)t}+\gamma_k^*\hat{c}_{-{\bf k}}^{\dagger}e^{i\omega(k)t}\right)e^{i{\bf k}{\bf r}}}{\sqrt{1-|\gamma_k|^2}},
\]
where $\mu=mc^2$, $\gamma_k=e^{-2i\Phi_0}(\hbar\omega(k)-\hbar^2k^2/2m-mc^2)/mc^2$,  $\hat{c}_{{\bf k}}$, $\hat{c}_{{\bf k}}^{\dagger}$ are the annihilation and creation Bose operators for excitations in liquid  $^4$He and $\omega(k)$ is given by:
\begin{equation}
\omega(k)=k\sqrt{c^2+\frac{\hbar^2 k^2}{4m^2}}.
\label{19} 
\end{equation}
Here $c=\sqrt{G\rho_{s0}}/m$ is the velocity of sound in He II, $\rho_{s0}=\rho[1-(T/T_\lambda)^{5.6}]$ is the density of the superfluid component of the liquid $^4$He 
assuming thermodynamic equilibrium \cite{Huang}, and $T_\lambda=2.18~{\rm K}$ is the critical temperature. As an example for $T=1.1~{\rm K}$, $c=2.4\cdot 10^4~{\rm cm/s}$. The Bogoliubov's type spectrum in Eq. (\ref{19}) for acoustic waves excited around  the constant density $\rho_{s0}$ at wavenumbers $k<k_0$ 
is close to the linear spectrum $\omega(k)=ck$ coinciding with the experimental data for the phonon spectrum in He II \cite{Gib} when $T\ll T_\lambda$ ($\rho_n\ll \rho_s$).
This solution for the field operator $\hat{\psi}_n({\bf r},t)$ yields the effective Hamiltonian in the diagonal form $\hat{H}=E_0+\sum_{{\bf k}}\hbar\omega(k)\hat{c}_{{\bf k}}^{\dagger}\hat{c}_{{\bf k}}$ which at wavenumbers $|{\bf k}|<k_0$ describes the phonon excitations.

The Eq. (\ref{17}) can be written in typical hydrodynamic form in terms of the superfluid density $\rho_s$ and velocity ${\bf v_s}$ by standard definitions \cite{Pit}:
$\psi_s=\sqrt{\rho_s/m}\exp(i\Phi)$~and ${\bf v_s}=(\hbar/m)\nabla\Phi$. 
Using the characteristic size $d$ between parallel plates ($1{\rm D}$ geometry) and the flow velocity $v$, new dimensionless variables can be defined by $t'=vt/d$, ${\bf r}'={\bf r}/d$, ${\bf v'_s}={\bf v_s}/v$, $\rho_s'=G\rho_s/m^2v^2$,  $\rho_n'=G\rho_n/m^2v^2$ and the dimensionless hydrodynamic form of Eq. (\ref{17}) is
\begin{equation}
\frac{\partial{\bf v'_s}}{\partial t'}+({\bf v'_s}\cdot\nabla'){\bf v'_s}+\nabla'\rho'_s=\frac{1}{8\pi^2{\cal R}^2}\nabla'\left(\frac{1}{\sqrt{\rho'_s}}\nabla'^2\sqrt{\rho'_s}\right)-2\nabla'\rho'_n,
\label{20} 
\end{equation}
\begin{equation}
\frac{\partial \rho'_s}{\partial t'}+\nabla'(\rho'_s{\bf v'_s})=0.
\label{21} 
\end{equation}
Here ${\cal R}=mvd/2\pi\hbar$ is the dimensionless number characterizing the flow of the superfluid $^4$He in the channel. Using Eq. (\ref{18}) one obtains further equations for $\rho'_n$. These are found also to depend only on ${\cal R}$ and no other parameters. Thus ${\cal R}$, the ``quantum Reynolds number" is the unique dimensionless variable parameter completely characterizing the flow. We note that if both terms on the right hand side of Eq. (\ref{20}) are negligible (Thomas-Fermi approximation) then we get the classical Euler equation for potential flow of a nonviscous fluid with pressure $P=G\rho_s^2/2m^2$.
For large numbers ${\cal R}$ the quantum pressure term \cite{Putt} which scales as ${\cal R}^{-2}$ in Eq. (\ref{20}) becomes less important in comparison with the non-linear inertial term $({\bf v'_s}\cdot\nabla'){\bf v'_s}$ and the normal fluid fluctuation term $-2\nabla'\rho'_n$. In this case, as it is observed experimentally, for some large ${\cal R}$ (at $d=1~{\rm cm}$ and $v=1~{\rm cm~s}^{-1}$) the laminar flow becomes increasingly unstable leading eventually to turbulence \cite{Putt} . Because the type of flow depends only on the unique dimensionless number ${\cal R}$ one may conclude that the quantum super-flow turbulence (${\rm QSFT}$) arises when ${\cal R}>{\cal R}_c$ where ${\cal R}_c$ is critical number. This condition yields the ${\rm QSFT}$ at $v>v_c$ where the critical velocity is
\begin{equation}
v_c=\frac{2\pi\hbar{\cal R}_c}{md}.
\label{22} 
\end{equation}
The critical distance $d_c$ is the boundary point on the curves given by Eq. (\ref{15}) and Eq. (\ref{22}), which yields the critical number ${\cal R}_c$ as
\begin{equation}
{\cal R}_c =\frac{m\Gamma d_c^{3/4}}{2\pi\hbar}.
\label{23}
\end{equation}
The experimental value \cite{Putt} of the critical width is $d_c= 1~{\rm cm}$ and hence from Eq. (\ref{23}) it follows that ${\cal R}_c\simeq1000$. This large number means that the de Broglie wavelength $\lambda$ of the helium atoms in the channel flow is much less then the channel size $d$ for ${\cal R}={\cal R}_c$. Thus the ${\rm QSFT}$ is the quasi-classical regime because the limit ${\cal R}\rightarrow \infty$ formally is equivalent to the limit $\hbar\rightarrow 0$. However we note that this regime arises from fluctuations of the normal component (see Eq. (\ref{18})) and has a quantum nature because ${\cal R}$ is finite. Fig. 1 shows the transition from superfluidity to ${\rm QRT}$ and ${\rm QSFT}$ regimes as predicted by Eq. (\ref{15}) for $d<1$\,cm ($v_c=d^{-1/4}$ c.g.s.) and Eq. (\ref{22}) for $d>1$\,cm ($v_c=d^{-1}$ c.g.s.) respectively.

In conclusion, in this paper are considered two different mechanisms of transitions of the superfluid $^4$He for $1{\rm D}$ geometry to quantum turbulence regimes when the influence of the normal fluid on superfluid flow is suppressed by superleaks. The first mechanism takes place for thicknesses $0.8\cdot 10^{-6}~{\rm cm}<d<1~{\rm cm}$ and $T<2~{\rm K}$, and can be explained by excitation of the rotational states of the rotons in the vicinity of the surface of the channel and further exchange of the rotational energy between roton clusters. In this case the functional form of the critical velocity is given by Eq. (\ref{15}) and matches the observed trend in experiment. The second mechanism of transition to quantum turbulence arises at $d>1~{\rm cm}$ and ${\cal R}={\cal R}_c\simeq1000$ when the quantum pressure term is diminished and the laminar flow becomes unstable. In this case the functional form of the critical velocity is given by Eq. (\ref{22})--- a prediction whose validity (of interest) requires the extension of existing experimental data.

I am grateful to Dr M. J. Collett and particularly to Dr D. Wardle for numerous valuable comments and useful discussions of this work.

\end{document}